# Observing parity-time symmetry in diffusive systems

Pei-Chao Cao[1,2,3,4,5], Ran Ju[2,3,4,5], Dong Wang[2,3,4,5], Minghong Qi[2,3,4,5], Yun-Kai Liu[1], Yu-Gui Peng[1,†], Hongsheng Chen[2,3,4,5,†], Xue-Feng Zhu[1,†], Ying Li[2,3,4,5,†]

[1]*School of Physics and Innovation Institute, Huazhong University of Science and Technology, Wuhan 430074, China*

[2]*Interdisciplinary Center for Quantum Information, State Key Laboratory of Extreme Photonics and Instrumentation, ZJU-Hangzhou Global Scientific and Technological Innovation Center, Zhejiang University, Hangzhou 310027, China*

[3]*International Joint Innovation Center, The Electromagnetics Academy at Zhejiang University, Zhejiang University, Haining 314400, China*

[4]*Key Lab. of Advanced Micro/Nano Electronic Devices & Smart Systems of Zhejiang, Jinhua Institute of Zhejiang University, Zhejiang University, Jinhua 321099, China*

[5]*Shaoxing Institute of Zhejiang University, Zhejiang University, Shaoxing 312000, China*

†Corresponding authors: ygpeng@hust.edu.cn (Y. -G. P.); hansomchen@zju.edu.cn (H. Chen.); xfzhu@hust.edu.cn (X.-F. Z.) eleying@zju.edu.cn (Y. L.);

**Abstract** Phase modulation has scarcely been mentioned in diffusive systems since the diffusion process does not carry momentum like waves. Recently, the non-Hermitian physics provides a new perspective for understanding diffusion and shows prospects in the phase regulation of heat flow, for example, the discovery of anti-parity-time (APT) symmetry in diffusive systems. The precise control of thermal phase however remains elusive hitherto and can hardly be realized in APT-symmetric thermal systems due to the existence of phase oscillation. Here we construct the counterpart of APT-symmetric diffusive systems, *i.e.*, PT-symmetric diffusive systems, which can achieve complete suppression of thermal phase oscillation. We find the real coupling of diffusive fields can be readily established through a strong convective background, where the decay-rate detuning is enabled by thermal metamaterial design. Moreover, we observe phase transition of PT symmetry breaking in diffusive systems with the symmetry-determined amplitude distribution and phase regulation of coupled temperature fields. Our work uncovers the existence of PT-symmetry in dissipative energy exchanges and provides a unique approach for harnessing the mass transfer of particles, the wave propagation in strongly scattering systems as well as thermal conduction.

***Introduction*** Metamaterials for exotic wave manipulation produced instant impetus to the rapid advancements in thermal metamaterials for harnessing the energy diffusion[1-4]. For example, the coordinate transformation theory and the scattering theory inspired the concepts of cloaking[5-8], camouflage[9,10], and even coherent perfect absorption[11] in thermal diffusion. However, most of these studies focused on the arrangement of nature materials with limited thermal conductivities. It is thus crucial to introduce convective modulation as a new degree of freedom to enrich thermal regulation. Many approaches were proposed on the convection-enhanced thermal conduction[12-14] and convection-driven nonreciprocal heat transfer[15-19]. Since the heat flow cannot carry momentum, the physical processes in conductive and convective diffusive systems have hybrid features of wave-like propagation and dissipative diffusion[16,17]. In the Fourier's equation for heat conduction, the convective modulation corresponds to the first-order derivative term of temperature field, inevitably bringing fluctuation in heat transfer. Recently, the effective Hamiltonian approach has been successfully introduced into diffusive systems, for an insightful description of heat exchange and temperature field evolution[20]. For a coupled dimer with only dissipative thermal conduction, the Hamiltonian governing the diffusion process takes an anti-Hermitian form in comparison with the Hamiltonian of wave systems ($H_{\text{diffusion}}=iH_{\text{wave}}$), where the onsite decoupled diffusion rates and offsite coupling strength are both imaginary. Under this perspective, the complex band theory based on impaginary coupling has attracted widespread attention[21–23], greatly advancing our understanding of diffusive systems.

The phase concept was introduced in diffusion in the study of convection-driven heat conduction[20]. By adding onsite convections to pure thermal diffusion, anti-parity-time (APT) symmetric thermal diffusion ($\{H, PT\} = 0$) can be established readily. The original pure imaginary eigenfrequencies of the effective Hamiltonian will turn into complex ones of real parts after passing through the exceptional points (EPs), where phase fluctuation occurs with the 1st-order relation to convection strength in the APT symmetry broken phase. As a result, the phase lag of coupled temperature fields was observed to change from stable to oscillating, leading to the important findings of EP-related thermal effects and topological heat transfer[24–28]. For precise phase regulation of temperature fields, we need to suppress the unwanted phase oscillations. Since the phase oscillation in APT broken phase is caused by real potential carried by convections, we resort to diffusive PT symmetry ($[H, PT]=0$) for stable phase regulation, where the real potential is introduced by the offsite real couplings[29–32]. However, the real coupling in diffusion systems is not intuitive to implement due to the dissipation nature of heat transfer. Moreover, the gain doping in construction of the diffusive PT symmetry was regarded as impossible since the thermal conductivity cannot be negative. Thus, the establishment of diffssuive PT symmetry is quite challenging and still remains elusive.

In this work, we propose an approach to realize an effective real coupling for two temperature fields by introducing very strong background convections between the two fields. The convection layer can be regarded as providing an effective infinite thermal conductivity[13] between the two temperature fields, where the coupling process can thus be regarded as no disspation inside and the offsite real coupling is introduced indirectly. As a result, only phase evolution can be observed in the convection layer. The phase fluctuation factor $\Omega$ in the coupled diffusive systems, proportional to the convection strength, is transferred to effective real coupling ($h^2/\Omega$) between the two temperature fields. Note that $ih$ denotes the imaginary coupling strength between the two directly contacted tempreture fields. Although the negative thermal conductivity is unreachable in passive systems, we utilize dissipation detuning (thermal metamaterial designs) to realize the unbalanced imaginary potential modulations and realize diffusive PT symmetry. Our experiments show the existence of EP and PT phase transition in the heat transfer, where both amplitude and phase of temperature fields can be precisely controlled with the phase fluctuation completely suppressed. Our work enriches current understandings of thermal transfer from finding new symmetries in diffusive systems and is important in thermal phase regulation.

***Real coupling in diffusive systems***. In previous studies, the temperature field evolution in a diffusive system can be calculated by thermal eigenmode analysis, where the eigen-frequency is complex with the imaginary part denoting the decay rate and the real part of convection term denoting the phase fluctuation[20,24] (Supplementary Materials). For two coupled temperature fields, heat exchange coefficient is inherently imaginary and from the Newton's law of cooling is determined by the material (or structural) parameters. However, when the convection carrying real potential dominates the physical process, this imaginary coupling can possibly be transformed into a real one, which is shown schematically in Fig. 1**a**. Therefore, both the imaginary-coupling-based APT symmetry and the real-coupling-based PT symmetry can be constructed in diffusive systems under different convection strengths.

We can utilize the effective Hamiltonian to theoretically investigate the convection-induced real coupling and PT symmetry. As shown in Fig. 1**b**, two temperature fields $T_1$ and $T_2$ are coupled through a convective middle layer at $T_3$ with the rotational angular velocity $\Omega$. According to the Fourier's law, the three-level thermal coupling equation can be expressed as

$$\rho_1 C_{p1} \frac{\partial T_1}{\partial t} = \kappa_1 \frac{\partial^2 T_1}{\partial x^2} + h_1(T_3 - T_1),$$
$$\rho_2 C_{p2} \frac{\partial T_2}{\partial t} = \kappa_2 \frac{\partial^2 T_2}{\partial x^2} + h_2(T_3 - T_2), \quad (1)$$
$$\rho_3 C_{p3} \frac{\partial T_3}{\partial t} = \kappa_3 \frac{\partial^2 T_3}{\partial x^2} - \rho_3 C_{p3} v \frac{\partial T_3}{\partial x} + h_3(T_1 + T_2 - 2T_3),$$

where $\rho_{1,2,3}$, $C_{p1,2,3}$, $T_{1,2,3}$, $\kappa_{1,2,3}$ and $h_{1,2,3}$ denote the mass density, specific heat, temperature field, thermal conductivity and heat exchange coefficient of channels 1, 2, 3, respectively. $v$ is the linear velocity of $T_3$. Temperature field in each channel takes the form of $T = T_0 + Ae^{i(kx-\omega t)}$, where $T_0$ is the reference temperature, $A$ is the amplitude of temperature modulation, $k$ is the propagation constant and $\omega$ is the eigenfrequency. Heat exchange coefficient can be deduced from the Newton's law of cooling, and the imaginary coupling strength is derived to be $ih_{1,2,3} = i\frac{\kappa_i}{\rho_{1,2,3}C_{p1,2,3}bd}$. Here, the coupling strengths can be regarded as $ih_0$ and the reference temperature is set to be zero. Since strong convection enhances thermal conduction extensively, the temperature field $T_3$ in the middle layer attenuates much faster than $T_1$ and $T_2$. The steady state is reached quickly with $T_3 = \frac{h_0}{(k^2D_3+2h_0+i\Omega)}(T_1 + T_2)$, where $k^2D_3$ is the decay rate of decoupled convectional background (Supplementary Materials). By substituting $T_3$ into Eq. (1), the three-level thermal system can be reduced to a dual-level one, where the imaginary coupling becomes complex as $\frac{h_0^2}{[\Omega-i(k^2D_3+2h_0)]}$, and even real ($h \approx \frac{h_0^2}{\Omega}$) as the convection strength dominates, i.e., $|\Omega| \gg k^2D_3 + 2h_0$. Obviously, the thermal phase fluctuation $\Omega$ is transferred to the real coupling of temperature fields. On the basis of real coupling, diffusive PT symmetry can thus be constructed by introducing different decay rates to the coupled temperature fields, similar to the case of passive PT-symmetric wave systems[33,34]. We note that in diffusive APT symmetric systems, the phase fluctuation $\Omega$ is imposed to the onsite temperature fields directly and the phase oscillation is obvious in the broken phase. However, in diffusive PT symmetric systems, the phase fluctuation $\Omega$ is introduced indirectly to the relatively weak real coupling. Therefore, the onsite phase oscillation can be largely suppressed in both PT symmetric and PT symmetry-broken phases, as shown in Fig. 1c, which is beneficial for the precise control of spatial phases of temperature fields.

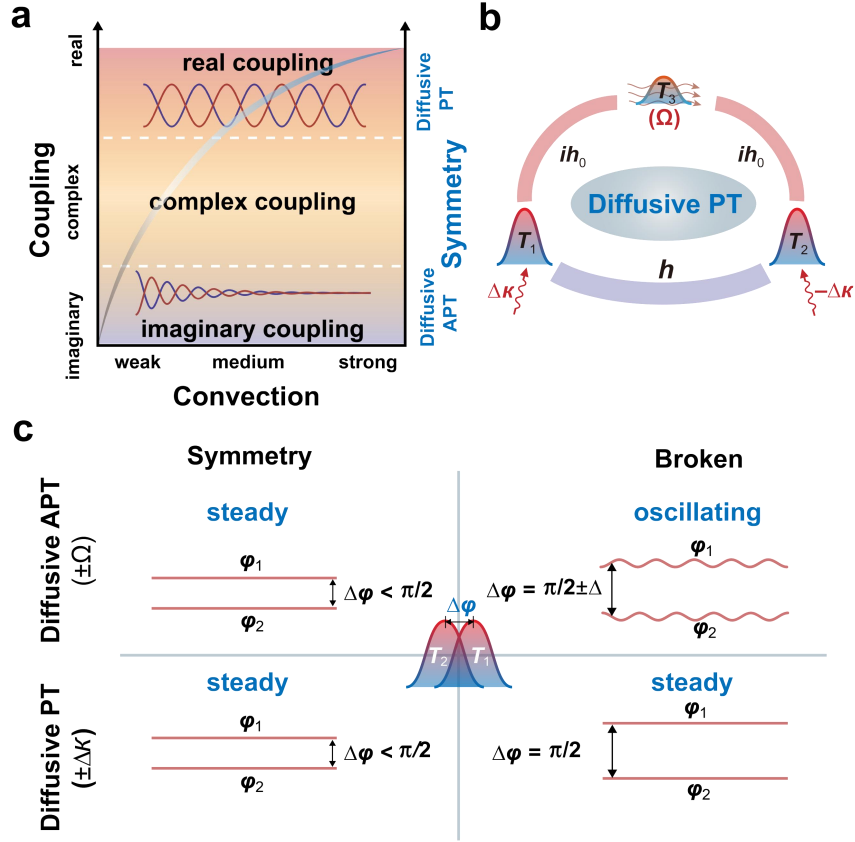

**Fig. 1. PT symmetry in real-coupled diffusive systems. a,** A schematic of convection strength dependent thermal coupling that evolves from imaginary to real when the convection strength increases. The APT and PT symmetry in diffusive systems can be constructed via the imaginary and real coupling, respectively. **b,** Diffusive PT symmetry based on effective real coupling. By introducing a strong convective layer ($T_3$, $\Omega$) between two temperature fields ($T_1$ and $T_2$), the direct imaginary coupling $ih_0$ can construct an indirect real coupling of $h$. The onsite decay rate detuning can be added through different thermal conductivities. **c,** A schematic of thermal phase regulation in APT/PT symmetric/broken diffusive systems. Note that phase oscillation in PT thermal systems are completely suppressed, providing the potential of precise thermal phase manipulation.

*Diffusive PT symmetry.* We show here the phase diagram of APT/PT symmetry/broken in Fig. 2**a**. The three-level system is still featured with imaginary coupling in the weak convection regime, where diffusive APT symmetry/broken can be explored under this perspective (Supplementary Materials). Effective real coupling comes into being in the strong convection regime, where diffusive PT symmetry/broken can be observed. Note that there exists a complex-coupling region in which the APT symmetry broken and PT symmetry phases are mixed[35]. Eq. (1) can be further represented by effective Hamiltonian and the temperature field evolution is transformed into the eigen problem of $H_{eff}$. The dual-level $H_{eff}$ with real coupling is derived as

$$H_{eff} = h\hat{\sigma}_x + i\delta\hat{\sigma}_z + (h - iS_0)\hat{I}, \qquad (2)$$

where $\hat{\sigma}_x$ and $\hat{\sigma}_z$ are the Pauli matrices, $h = h_0^2/\Omega$ is the real coupling strength, $\delta = \frac{1}{2}\frac{k^2}{\rho c_p}(\kappa_1 - \kappa_2) + h_0$ and $S_0 = \frac{1}{2}\frac{k^2}{\rho c_p}(\kappa_1 + \kappa_2) + h_0$ are the detuning and average of decay rates, respectively. It should be mentioned that the imaginary coupling also contributes to the decay rates in thermal diffusion. Therefore, the decay rate detuning can also be affected by the asymmetric imaginary couplings of temperature fields[21,28,36]. The details of asymmetric-imaginary-coupling induced PT symmetry are attached in Supplementary Materials. For simplicity, we focus on the symmetric-coupling case here, for which the eigenfrequency of the effective Hamiltonian can be deduced as

$$\tilde{\omega}_{1,2} = \pm \sqrt{h^2 - \delta^2}, \qquad (3)$$

where $\tilde{\omega}_{1,2} = \omega_{1,2} - (h - iS_0)$ with $h - iS_0$ the reduction term. Eigenfrequency of the convective middle layer is $\tilde{\omega}_3 = \Omega$. Obviously from Eq. (3), the eigenfrequencies degenerate at $h = \delta$, where the real coupling strength equals to the decay rate detuning. Before this EP, the convection is weak with $h > \delta$ and $\tilde{\omega}_{1,2}$ is real. After exceeding the EP with the convection enhanced, we have $h < \delta$ and $\tilde{\omega}_{1,2}$ becomes imaginary. As shown in Fig. 2b, the EP is actually the self-bifurcation singularity point of both eigenstates and eigenfrequencies of Rieman surfaces in non-Hermitian parameter space.

The nonorthogonal eigenvectors of Eq. (2), *i.e.*, eigenstates of temperature fields, are solved to be $\Psi_{1,2} = (\pm e^{\pm i\theta}, 1)^{\mathrm{T}}$ for the diffusive PT symmetry phase ($\theta =$ $a\sin\frac{\delta}{h}$). The eigenstates $\Psi_{1,2}$ can go back to itself after a PT operation. For the diffusive PT symmetry-broken phase, $\Psi_{1,2} = (ie^{\pm\emptyset}, 1)^T$ with $\emptyset = \mathrm{acosh}\frac{\delta}{h}$ and $PT\Psi_{1,2} \neq \Psi_{1,2}$. We find that in the PT symmetry phase the coupled temperature fields always keep a steady phase lag $\theta$ and the same amplitude. However in PT symmetry-broken phase, the temperature fields have an unchanged $\pi/2$ phase lag and unbalanced amplitudes. This result is shown in the eigenstate simulations with three channels that satisfy periodic boundary conditions, as shown in Figs. 2c and 2d. Note that in Fig. 2d the temperature field $T_3$ in the convective middle layer fades away quickly to form a dark state, where the difference $(T_{3,\max} - T_{\mathrm{ref}})$ is small and can be ignored.

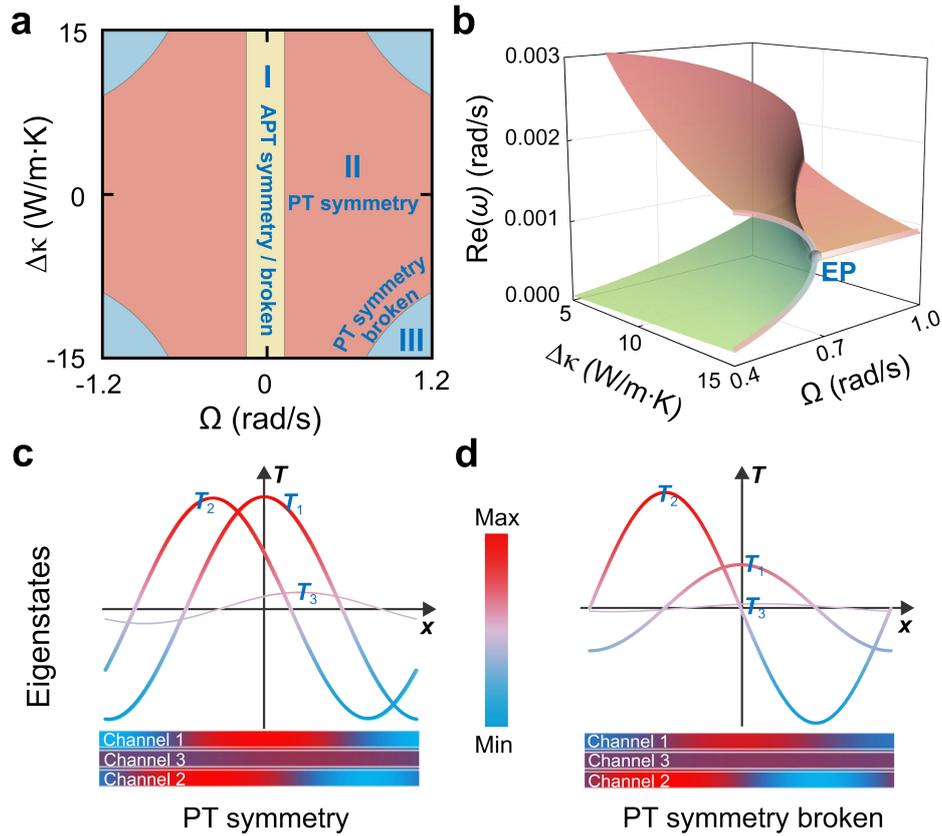

**Fig. 2. Diffusive PT symmetry and EP. a**, The phase diagram of APT/PT symmetry/broken. The diffusive system is in the APT symmetry/broken phase (**I**) when the convection strength is weak and comparable with the imaginary coupling. When the convection strength increases and dominates, the system turns into PT symmetry phase (**II**), and finally exceeds the EP and breaks the PT symmetry (**III**). **b**, Real part of eigenfrequency of the diffusive system versus

decay rate detuning and convection strength. Note that the upper and lower branches merge at an EP. **c** and **d**, Eigenstates at the diffusive PT symmetric and symmetry-broken phases, where phase lag is $\Delta\varphi < \pi/2$ (or $\Delta\varphi = \pi/2$) and amplitude is uniform (or unbalanced) in PT symmetry (or symmetry-broken) phases.

***Design of diffusive PT systems***. To implement the diffusive PT systems, we utilize the coupled metamaterial ring structures, for which the temperature fields are loaded and the convective background is mimicked by ring rotation. Since the thermal gain is very challenging with the requirement of negative thermal conductivity, we take the passive PT approach instead via effective thermal decay rate detuning[33,34]. The thermal decay rate detuning can be readily introduced by different thermal metamaterial designs. However, this will inevitably induce asymmetric couplings of temperature fields due to different mass densities[21,28,36], and the non-Hermitian potential will make the temperature field evolution very complicated (Supplementary Materials). In order to realize the thermal decay rate detuning without changing the mass density, here we adopt the anisotropic thermal metamaterial design, for which effective thermal conductivity can be obtained by the Garnett-Maxwell's equation. As shown in Fig. 3**a**, the through holes filled with air (low thermal conductivity) are uniformly distributed on the copper ring (high thermal conductivity) to adjust the effective thermal conductivity into a lower level. For the anisotropic design, the slit array is radially distributed on the copper ring to adjust the effective conductivity into a different level. In both cases, the effective mass densities $\rho_{C1,2}$ keep equivalent by setting the same filling ratio of air. In Fig. 3**b**, we present the decay rate detuning via transient simulations for the two different ring structures in Fig. 3**a**. Here we excite the first-order polarization temperature eigenfield (inset figure). The attenuation curve of temperature fields can be well described by $e^{-k^2 \frac{\kappa_{1,2}}{\rho c_p} t}$, where $k = 1/R$ denotes the propagation constant and $R$ is the ring radius. Figure 3**b** shows the existence of decay rate detuning, where the temperature attenuation in the isotropic metamaterial is faster than the one in the anisotropic metamaterial.

In Fig. 3c, we show the characteristics of diffusive PT symmetry in our established model. The eigenfield simulation shows that the phases of temperature fields $T_1$ and $T_2$ together with the phase lag are all steady in the final states, even though the background has a nontrivial convection. When the convection reaches an EP ($\Omega = 1.20$ rad/s), the phase lag increases to $\pi/2$. The $\pi/2$ phase lag holds after the convection is above the EP ($\Omega > 1.20$ rad/s). Note that the normalized intensities of temperature modulations are steady in both diffusive PT symmetry and symmetry-broken phases. Here we define the normalized intensity by $A^2_{\text{norm}1,2} = A^2_{1,2}/(A^2_1 + A^2_2)$. Specifically, in PT symmetry phase, the intensities in the two coupled channels are almost the same. However, in the PT symmetry-broken phase, the intensities are very different in the two channels due to the non-unitary eigenvectors. More analyses can be found in Supplementary Materials, where the thermal APT case is also discussed as a contrast.

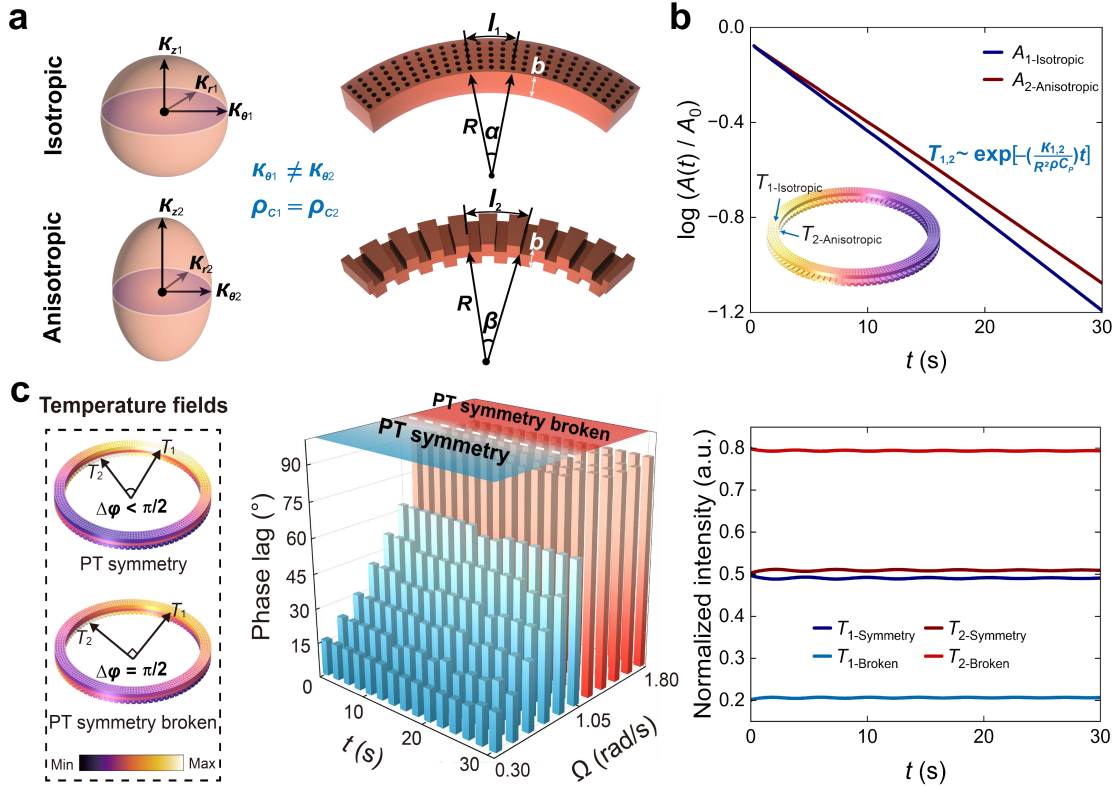

**Fig. 3. Design of diffusive PT symmetry systems. a,** Thermal metamaterial design for decay rate detuning. Isotropic and anisotropic thermal metamaterials can be realized by the through-hole array and surface-groove array, respectively. **b,** Transient simulation of temperature field evolutions in the meta-rings with isotropic and anisotropic thermal conductivities, respectively. **c,** Simulation of temperature eigenfields of diffusive PT symmetric systems. The convective background is driven by rotating the rings. In the PT symmetric phase, temperature fields in the coupled rings have a steady phase lag ($< \pi/2$) as well as the balanced temperature distributions. However, in the PT symmetry-broken phase, the phase lag keeps at $\pi/2$ and the temperature distributions are unbalanced.

*Experimental observation*. In our experiments, the isotropic and anisotropic meta-ring channels were fabricated with red copper ($\kappa = 394 \text{ W/m·K}$, $\rho = 8900 \text{ kg/m}^3$, $C_p = 385 \text{ J/(kg·K)}$). The convective middle ring was fabricated with aluminum alloy ($\kappa = 180 \text{ W/m·K}$, $\rho = 2700 \text{ kg/m}^3$, $C_p = 880 \text{ J/(kg·K)}$). Experimental demonstration of thermal decay rate detuning is attached in Supplementary Materials. The schematic of our setup is shown in Fig. 4**a**, where we excited eigenfields by loading temperature fields on isotropic ring 1 and anisotropic ring 2 with one side hot (323 K) and the other side cold (263 K). Three independent motors were utilized to drive the meta-rings for creating a convective background and offseting the global phase shift induced from the convection. To reduce thermal interface resistance and obtain uniform emmisivity, we used the silicone grease (2 W/m·K) and silicone oil (0.16 W/m·K). All supporting parts were nylon (4 W/m·K) for suppressing the radial temperature field diffusion.

In Fig. 4**b**, we show the results of the thermal phase oscillation suppression. In the experiments, we set the rotational speeds of motors to be $\Omega_1 = -0.035 \text{ rad/s}$, $\Omega_2 = -0.035 \text{ rad/s}$, and $\Omega_3 = 0.85 \text{ rad/s}$, respectively. The initial temperature fields are eigenstate excitation, with the same temperature difference on the two channels and the phase lag between $T_2$ and $T_1$ being $\pi/4$. Temperature fields on the two meta-rings were recorded by two infrared cameras (Fotric 628). The measurement areas are marked by the white dashed lines in the inset of Fig. 4**b**. Our measured results unambiguously show that the phases and normalized amplitudes of temperature fields on the two rings are very steady in the time window of $t_0 \rightarrow$ $t_1$: 15 s, with only $\pi/40$ phase shift in the half channels. It should be mentioned that we used consine-fitted curves to capture the positions of $T_{\max}$ for characterizing the thermal phase. The even mode excitation case is presented in Supplementary Materials for a comparison.

We furtherly show the observation of diffusive PT symmetry and symmery-broken phases. The experimental results are consistent with the theoretical ones, as shown in Figs. 4c-d, where in the PT symmetric phase the two coupled temperature fields have a phase lag of ($< \pi/2$) and balanced intensities (0.5, 0.5) and in the PT symmetry-broken phase the two fields have a phase lag locked at $\pi/2$ and unbalanced intensities ($0.5 - \Delta A^2, 0.5 + \Delta A^2$). In Fig. 4e, we show the final states of temperature distributions in the PT symmetry and symmetry-broken phases under different convection strengths, where the phase lag is marked by the arrows that point to the temperature maximum $T_{\max}$. Details of analyses are shown in Supplementary Materials.

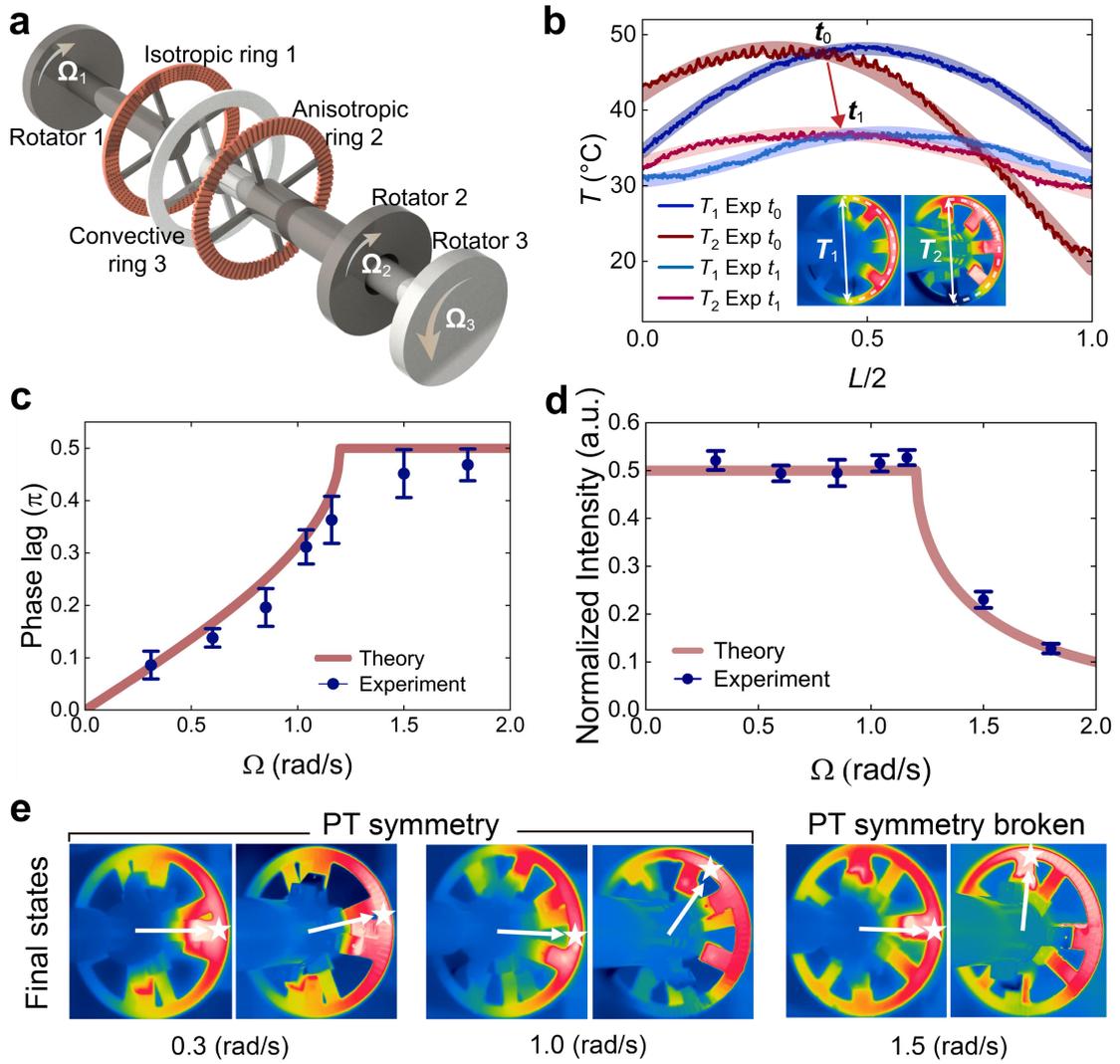

**Fig. 4. Observation of diffusive PT symmetry. a**, The experimental setup. The isotropic and anisotropic metamaterial rings as well as the convective middle layer are fabricated by copper and aluminum alloy, respectively, which are driven by motors with different rotational speeds $\Omega_{1,2,3}$. **b**, Suppression of temperature phase oscillation at a strong convection of 0.86 rad/s. Note that the spatial phases of temperature fields are almost unchanged. **c** and **d**, Phase lag and intensity ratio of the coupled temperature fields under the diffusive PT symmetry/broken phases. **e**, Final states of the two temperature fields with eigenstate excitations at diffusive PT symmetry and broken phases.

***Discussion***. In this work, we present the experimental observation of the PT symmetry and symmetry-breaking in diffusive paradigm. In the construction of PT symmetry, we realize effective real couplings for two interacting diffusive fields by introducing a strong convective backgound. Inspired from previous passive PT symmetry approach in wave systems, we adopt the thermal metamaterial design to realize the decay rate detuning. We prove the existence of PT symmetry breaking in diffusive thermal systems, where the phase and amplitude of temperature fields are experiencing the changes from nonorthogonal to orthogonal and from balanced to unbalanced, respectively. Due to the fact that the nontrivial convection is transformed into the weak real coupling term, the onsite thermal phase oscillation is siginficantly suppressed, enabling a precise control of spatial phases of temperature fields. Our work changes the understanding of energy exchange and thermal management in diffusive systems from symmetry-induced phase transition perspective, which can be useful for high-sensitivity thermal detection[37] and robust thermal meta-device designs[38]. The expansion of PT symmetry into the diffusive regime can arouse interests in the research fields of optics, condensed matter physics as well as quantum mechanics.

**Data availability**

The data that support the findings of this study are available from the corresponding authors upon request.

**Code availability**

The code utilized during the current study is available from the corresponding author upon request.

**Acknowledgements**

The work is sponsored by the Key Research and Development Program of the Ministry of Science and Technology under Grant 2022YFA1405200, the National

Natural Science Foundation of China under Grants Nos. 92163123 and 52250191, the China Postdoctoral Science Foundation under Grant 2023M733120, the Key Research and Development Program of Zhejiang Province under Grant No. 2022C01036.

**Author contributions**

Pei-Chao Cao, Ying Li and Xue-Feng Zhu conceptualized this work and carried out the theoretical modelling, simulations and data analyses. Ran Ju, Minghong Qi, Dong Wang and Yun-Kai Liu provided important suggestions. Ying Li, Xue-Feng Zhu, Hongsheng Chen and Yu-Gui Peng supervised this project. All authors wrote the paper.

*Rev. Mod. Phys.* **93**, (2021).